\documentclass[pre,twocolumn,longbibliography,amsmath,amssymb,floatfix]{revtex4-1}

\usepackage{graphicx}
\usepackage{dcolumn}
\usepackage{bm}
\usepackage[usenames,dvipsnames]{xcolor}
\usepackage{mciteplus}
\usepackage{braket}

\newcommand{\uref}{{\bf u}_{\mathrm{ref}}}
\newcommand{\Oref}{\Omega_{\mathrm{ref}}}
\newcommand{\Sref}{\mathcal{S}_{\mathrm{ref}}}
\newcommand{\bu}{{\bf u}}
\newcommand{\bx}{{\bf x}}
\newcommand{\bk}{{\bf k}}

\def\ADDB#1{{\textcolor{magenta}{#1}}}      % addition (Luca)
\begin{document}
\title{Inferring flow parameters and turbulent configuration with physics-informed data-assimilation and spectral nudging}\thanks{Postprint version of the manuscript
  published in Phys. Rev. Fluids {\bf 3}, 104604 (2018).}
\author{Patricio Clark Di Leoni$^{1}$\email{patricio.clark@roma2.infn.it},
  Andrea Mazzino$^2$\email{andrea.mazzino@unige.it}, \&
  Luca Biferale$^1$\email{luca.biferale@roma2.infn.it}}
\affiliation{$^1$Department of Physics and INFN, University of
    Rome Tor Vergata, Via della Ricerca Scientifica 1, 00133 Rome, Italy.\\
                  $^2$Department of Civil, Chemical, and Environmental
                  Engineering and INFN, University of Genova, Genova 16145,
                  Italy.}
\date{\today}
\begin{abstract}
Inferring  physical  parameters of turbulent flows by assimilation of
data measurements  is an open   challenge with key applications in
meteorology, climate modeling and   astrophysics.  Up to now, spectral
nudging was applied for empirical data-assimilation as a mean to improve
deterministic and statistical predictability in the presence of a
restricted set of field measurements only.  Here, we explore under which
conditions a nudging protocol can be used for two novel objectives: to
unravel the value of the physical flow parameters  and to reconstruct
large-scale turbulent properties starting from a sparse set of
information in space and in time.  First, we apply nudging to
quantitatively infer the unknown rotation rate and the shear mechanism
for turbulent flows. Second, we show that a suitable spectral nudging is
able to  reconstruct the energy containing scales in rotating turbulence
by using a blind set-up, i.e.  without any input about the external
forcing   mechanisms acting on the flow. Finally,  we discuss the broad
potentialities of nudging to other key applications for physics-informed
data-assimilation in  environmental or applied  flow configurations.
\end{abstract}
\maketitle

% Introduction 
\section*{Introduction} 

Extracting information from experimental or observational data
of fluid flows is a highly challenging task.  While in laboratory
experiments one can control and/or measure the properties of the system
(e.g. viscosity, thermal expansion coefficient, large scale shear,
rotation rate etc...), this is  often impossible when performing
observations in the open field, such as for meteorological data taken
from the atmosphere or astrophysical data in the sky.  Thus, one has to
resort to other methods to infer  the desired parameters, a task which
most of the time is obstructed by the quality of the data at hand.  The
problem is part of a vaster paradigm that goes under the name of data
assimilation and optimal reconstruction, where one is faced with the
need to infer the flow parameters or to extrapolate measurements from a
sparse sub-volume of the flow field to the whole space.  The problem is
also connected to the need to control and improve predictability
for the evolution of chaotic systems by using only a partial set of
information about the full trajectory.  These problems can be
encountered in a wide range of fields, going from atmospherics sciences
\cite{Akyildiz02,Hart06}, astrophysics \cite{Fu15}, optics
\cite{Carpeggiani17} and medical physics \cite{Busch13}.  Several tools
have been developed to tackle these challenges. In the context of
numerical weather prediction, variational principles and ensemble
filters have been developed to fine-tune the parameters entering in the
sub-grid models \cite{Kalnay,Anderson99,Anderson01,Ruiz13}.
Alternatively, other techniques coupled with Bayesian inference, machine
learning and deep learning have been proposed to estimate the parameters
phase-space in Reynolds-averaged Navier-Stokes models in engineering
problems \cite{Kennedy01,Xiao16,Parish16,Ling16}. Also, information
theory and statistical mechanics tools such as belief propagation have
been used to infer parameters from turbulent flows by looking at the
motions of transported particles \cite{Chertkov10}. Another interesting
example is the use of sparse regression methods to discover not only
parameters but the actual form of the terms controlling the evolution of
a system \cite{Brunton16,Rudy17}.

In this paper, we explore a new avenue and we show how to infer the
physical flow parameters from partial data assimilation by exploiting
the equations of motion in a dynamical way, using a technique known as
{\it nudging}, whose conceptual foundation goes well beyond applications
to physics (see 2017 Nobel lecture on Economy by R.E.  Thaler). Contrary
to the attempts previously mentioned where the modeled flow is usually
compared with data by using a cost function, nudging introduces an extra
term in the dynamical equation where {\it partial} information from
field measurements is inputted and exploited to reconstruct the
unmeasured degrees-of-freedom.  Nudging, has been successfully used and
developed to input global circulation model into a regional climate
model \cite{Waldron96,Vonstorch00,Miguez-macho04}. In this case, due to
computational constrains, the global models can not solve the smallest
dynamically active  scales so as to have accurate local weather
predictions, while the regional models can not solve for the large
planetary cyclonic and anticyclonic circulations.  Nudging is applied
to match the overlapping scales in each model by forcing the regional
model to behave as the global one via a penalty term.  Outside numerical
weather prediction, nudging has also been rigorously applied to estimate
bounds in the data assimilation problem in two dimensional Navier-Stokes
equations \cite{Farhat16,Gesho16}, the three dimensional Navier-Stokes
$\alpha$-model \cite{Albanez16}, and in Rayleigh-Bernard convection
\cite{Farhat17}. It has also been used to study synchronization in maps
and dynamical systems \cite{Pazo14}. To the best of our knowledge, no attempts have ever
been made to benchmark and optimize its performances to the
three-dimensional Navier-Stokes equations in the fully developed
turbulent regime, characterized by high chaoticity and by a
high-dimensional strange attractor.

We  implement a spectral nudging technique with two novel aims.  First,
we show how to use  nudging  as a physics-informed tool to  accurately
infer key flow parameters as, e.g. the rotation rate or the large-scale
stirring mechanism, from a limited sub-set of data sparsely measured in
time and in Fourier space.  Second, we show that the same technique can
be used to {\it learn}  the global physical turbulent configuration. We
do this by using the nudged equations to reconstruct in space the
large-scale energy distribution of rotating turbulence under the
presence of a split energy cascade and without inputting in to the
algorithm any information about the external forcing mechanism and about
the intensity of the rotation rate. Nudging is thus presented as a
general data-driven algorithm to learn from sparse measurements in a
dynamical way and a with a broad range of applications. Finally, we
discuss a series of open challenges to adapt and extend the application
of nudging to other turbulent flow configurations using either Eulerian
or Lagrangian field measurements and in different domains.

%%%%%%%%%%%%%%%%%%%%%%%%%%%%%%%%%
\begin{figure}
    \centering
    \includegraphics[width=8.5cm]{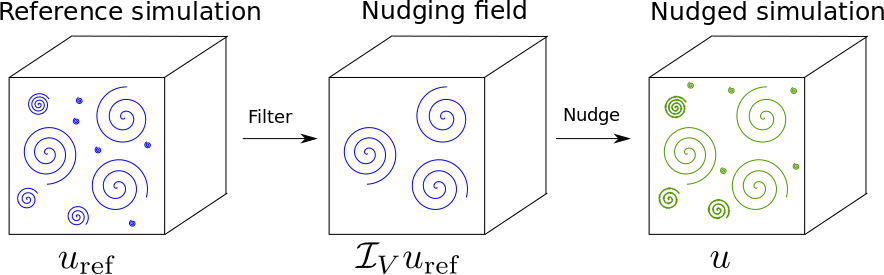}
    \caption{Diagram showing the set-up of our numerical experiments.
    First, a reference simulation is performed (left). Second,  a subset
    of data is filtered out of the reference field, by keeping only data
    on a given sub-set of points in space and instant in times (center).
    Third, we interpolate in time the input partial information and use
    it to nudge the evolution of a new field to reconstruct the missing
    data and to infer the correct physics parameters (right).}
    \label{diag}
\end{figure}
%%%%%%%%%%%%%%%%%%%%%%%%%%%%%%%%%%%%%%%%%%

\section*{The nudging technique}

As said,  nudging means to {\it gently convince}  a numerical flow to
evolve  as close as possible  to a reference set supposing to have only
partial measurements or observations of the latter
\cite{Waldron96,Vonstorch00,Miguez-macho04}. The idea is to use the
equation of motion to perform an optimal data and flow-parameter
assimilation in the interval of time $t \in (0,t')$ and in the whole
fluid volume.  Suppose we have a reference three dimensional turbulent
flow, $ {\bf u}_{\mathrm{ref}}({\bf x},t)$ evolving under the action of
a set of external forces, ${\bf {\cal F}}[{\bf u}_{\mathrm{ref}}, {\cal
V}_{\mathrm{ref}}]$, parametrised by a set of physical coefficients,
${\cal V}_{\mathrm{ref}} =(\Omega_{\mathrm{ref}}, {\cal
S}_{\mathrm{ref}}, \ell_{\mathrm{ref}}, \Delta T_{\mathrm{ref}},
\cdots)$ where we denoted with $\Omega_{\mathrm{ref}}$ the rotation
rate, with ${\cal S}_{\mathrm{ref}}$ the amplitude of a large scale
shear with the typical length scale $\ell_{\mathrm{ref}}$, with $\Delta
T_{\mathrm{ref}}$ the temperature difference across the volume etc...
Suppose that we have access to the measurements of the reference
velocity field, ${\bf u}_{\mathrm{ref}}$ on a limited set of  $M$
anemometers placed in  ${\bf x}_j$ with $j=1,\cdots,M$ that record the
flow properties at  $N$ time instants $t_n$ with $n=1,\cdots,N$ , i.e.
we control ${\bf u}_{\mathrm{ref}}$ in a given sub-domain of the whole
space-time (3+1) volume only. The idea behind nudging is to evolve  an
independent three dimensional incompressible Navier-Stokes (NS)
equations with an initially {\it educated guess} for the set of
parameters, ${\cal V}$, and imposing a penalisation whenever the flow
field does not reproduce the inputted velocity values of the reference
field in the space-time domain  $V =  ({\bf x}_j,t_n)$: 

\begin{equation}
    \frac{\partial {\bm u}}{\partial t}  + {\bm u} \cdot {\bm \nabla}
    {\bm u} =  - {\bm \nabla} p + \nu \nabla^2 {\bm u}  + {\bf {\cal
    F}}[{\bf u}, {\cal V}] - \alpha \mathcal{I}_{V}  ( {\bm u} - {\bm
    u}_\mathrm{ref})
    \label{nsnudged}
\end{equation}
where  $\nu$ is the viscosity, $p$ is the pressure that ensures the
incompressibility condition,  $\mathcal{I}_{V}$  is a
dimensionless linear projector
operator given by the characteristic function of the set ${V}$, and
$\alpha$ is a parameter that controls the intensity imposed by the
nudging {\it control} and has units of frequency.  In its crudest form, $\mathcal{I}_{V}$ is equal
to 1 for $(\bx,t) \in V$  and 0 otherwise. The simplest and most common
improvement is to linearly interpolate the different measured snapshots
between each time $t_n$ and $t_{n+1}$. So when entering
\eqref{nsnudged}, $\uref$ will always be assumed to be piece-wise
differentiable in time with a characteristic  interpolation window,
$\tau$. {In} this way the operator $\mathcal{I}_{V}$ is only acting on
the spatial part of the fields. The whole protocol is sketched in
Fig.~\ref{diag}.  It is important to realize that, in our application, we do not even require
to know the exact way the system is forced, i.e. we do not impose
${\cal V} = {\cal V}_{\mathrm{ref}}$ and  the only {\it a priori}
information that we provide is inside the partial measurements of the
reference field.  Clearly, the success of the reconstruction will depend
on the amount of information provided (how many measurements in space
and in time), on its {\it quality} (where and what we measure)  and  on
the intensity of the penalization term, $\alpha$. Notice that, because of potential stiffness and 
truncation effects arising when $\alpha$ is big,  it is not
{\it a priori} obvious that {taking} large $\alpha$ is {the best choice}.
 It is intuitive to imagine that in some cases it might be better to allow for
a larger  error in some measuring stations to allow the field to be
closer to the target globally.

\section*{Set-up of the numerical spectral nudging experiment}

We start first by restricting to the case when the set of {\it external}
parameters are given by the intensity of the  Coriolis force due to the
presence of a rotation $\Omega$ in the vertical direction and of an
external stirring mechanism $\bm{{\cal S}}$:

\begin{equation}
{\bf {\cal F}}[{\bf u}, {\cal V}]   = 2 \Omega \, \hat{{\bf z}} \times
\bu(\bx,t) + \bm{{\cal S}}(\bx).
\end{equation}
where $\bm{{\cal S}}$ is a randomly-generated, quenched in time,
isotropic field with support on wavenumbers with amplitudes $ k\in[k_{f1},k_{f2}]$ whose Fourier
coefficients are given by $\bm{\hat{{\cal S}}}({\bf k}) = \mathcal{S} k^{-7/2}
e^{i\theta_{\bf k}}$, where $\theta_{\bf k}$ are the random  phases.
In the remaining part of this paper we will address the most ideal case
when the information is supplied in Fourier space, i.e.  we imagine to
have a periodic array of measurement stations that allow us to
reconstruct the reference flow configuration in a given range of nudged
wavenumbers, $ k_0 < k  <k_1$.  In this case, the
$\mathcal{I}_{V}$ operator reduces to a band-pass Fourier filter of the
form  

\begin{equation}
    {\cal I}_{V} {\bf u} = \sum_{k_0 <   |{\bf k}| <k_1} {\hat
    {\bf u}}({\bf k},t) \exp{(i {\bf k}\cdot {\bf x})},
    \label{forcing}
\end{equation}
that projects the velocity field on the window of nudged Fourier modes.

We implement the whole protocol as follows. First we numerically produce
a full space-time evolution of the whole $\uref$ field in a interval $t
\in (0,T_{tot})$ by solving the Navier-Stokes equations with a
reference rotation rate $\Omega_{\mathrm{ref}}$ and a given intensity of
the shear ${\cal S}_{\mathrm{ref}}$ (i.e., Eqs~\eqref{nsnudged} with
$\alpha=0$). The values of $\Omega_{\mathrm{ref}}$ and ${\cal
S}_{\mathrm{ref}}$ (and also $\nu$ which is the same for both the
reference and the nudged simulations) are given in Table~\ref{table1}.
All reference simulations are started from rest and allowed to reach
stationary states ($t=0$ denotes the start of the stationary states).
Second, we extract the inputting field in a subset of discrete times
$t_n = n \tau$ with $\tau$  chosen as a fraction of the characteristic
eddy turnover time of the flow (see Table \ref{table1}) .  Third, we
define the nudging field (\ref{forcing}) by a linear interpolation
between $t_n$ and $t_{n+1}$ for all intervals.  The initial
condition used for all nudged simulation is just the first extracted
input field (i.e., the field at $t=t_0$) with all the modes outside the
nudging region filtered out. All simulations have been performed with a
parallel pseudo-spectral code.  The code uses a two step Adams Bashfort
scheme for the time integration, the ``2/3 rule'' for dealiasing and
periodic boundary conditions in all three directions.  In the following
we will analyze three different nudging protocols. The first two cases
are about  simulations made to {\it infer} the physical flow parameters,
$\Oref$ and $\Sref$ (called {\it INFER1} and {\it INFER2} in the
following, see also Table I for details). The third case is about the
reconstruction of the large-scale coherent structures and it is called
{\it PHYS1}.  Numerical details for all set-ups can be found   in
Table~\ref{table1}. The value of $\tau$ is such that it is smaller
than the decorrelation time of the fastest nudged mode, while $\alpha$
was taken as $1/\tau$, these choices follow common practices \cite{Omrani12}. A
comprehensive report about the performance of nudging at changing
$\alpha,\tau$ for fully developed homogeneous and isotropic turbulent
flow is not the scope of this paper and it will be presented elsewhere.

%could be controlled with this technique. Their values are $k_0=0$,
%$k_1=4$, and $\alpha=10$. While a proper explanation of how these
%%parameters affect the nudging in three-dimensional fully-developed
%turbulence is not simple and will be the subject of a future work what
%is important here is to understand that the protocol works in this
%setting, so choosing appropriate parameters is a matter of
%optimisation.

\begin{table}%[tbhp]
    \centering
    \begin{tabular}{c | c | c | c | c | c | c | c | c | c | c}
    Set-up & $E_{kin}$  & $T$ &
     $\nu$ & $\mathrm{Re}$ &
    $\Sref$ & $[k_{f1},k_{f2}]$ & $\Oref$ & $[k_0,k_1]$\\
    \colrule
    {\it INFER1} &  1.84 & 3.28 & 0.002  & 6030 &
    0.005 & [1,2] & 2 & $[1,4]$\\
    {\it INFER2} &  1.20 &  4.06 & 0.0025 &  4900 &
    0.02  & [1,2] & 0 & $[1,4]$\\
    {\it PHYS1} &  0.0012 &  128  & 0.002  &   150 &
    0.004 & [10,11] & 20 & $[8,20]$\\
    \colrule
    \end{tabular}
    \caption{Parameters used in the different numerical
    experiments.  {\it INFER1} is the set-up for the $\Omega$ scan, {\it
    INFER2} for the $\mathcal{S}$ scan, and {\it PHYS1} for the inverse
    cascade experiment. The values listed are the total kinetic energy
    $E_{kin}= 1/2 \langle |\bu|^2\rangle $, the eddy turnover time
    $T=L/(2 E_{kin})^{1/2}$, with $L=2\pi$ being the largest scale in
    the flow (same for all simulations), the viscosity $\nu$, the
    Reynolds number $\mathrm{Re}= L (2 E_{kin})^{1/2}/\nu$, the forcing
    intensity of the reference simulation $\Sref$, the band of forced
    wavenumbers in the reference simulation $[k_{f1},k_{f2}]$, the
    rotation frequency of the reference simulation $\Oref$, and the band
    of nudged wavenumbers $[k_0,k_1]$.  The number of grid points
    $N^3_{\mathrm{grid}}=256^3$, the time step of the simulations
    $dt=0.001$, the nudging intensity $\alpha=10$, and the temporal
    interpolation window of the nudging field $\tau=0.1$ are the same
    for all simulations.  The box length $L$, the temporal timestep
    $dt$, the resolution $N^3_{\mathrm{grid}}$, and the viscosity $\nu$
    are the same for the reference and the nudged simulations in each
    set. The kinetic energy and Reynolds numbers are given for the
    reference run of each set, the nudged ones have very similar
    values.
}
    \label{table1}
\end{table}

\section*{Inferring physical parameters in rotating turbulence}

%%%%%%%%%%%%%%%%%%%%%%%%%%%%%%%
\begin{figure}
    \centering
    \includegraphics[width=0.44\textwidth]{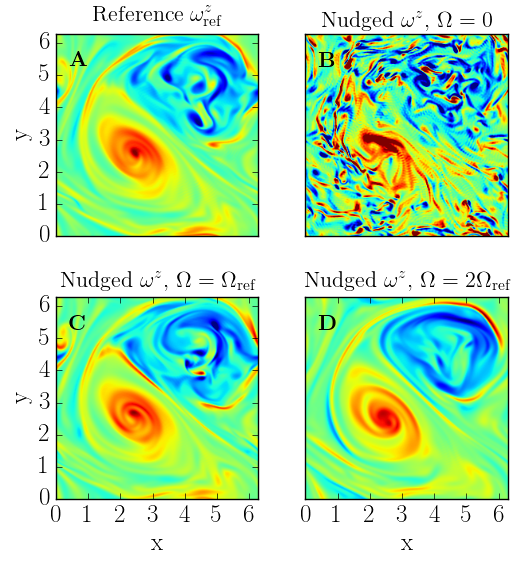}
    \includegraphics[width=0.44\textwidth]{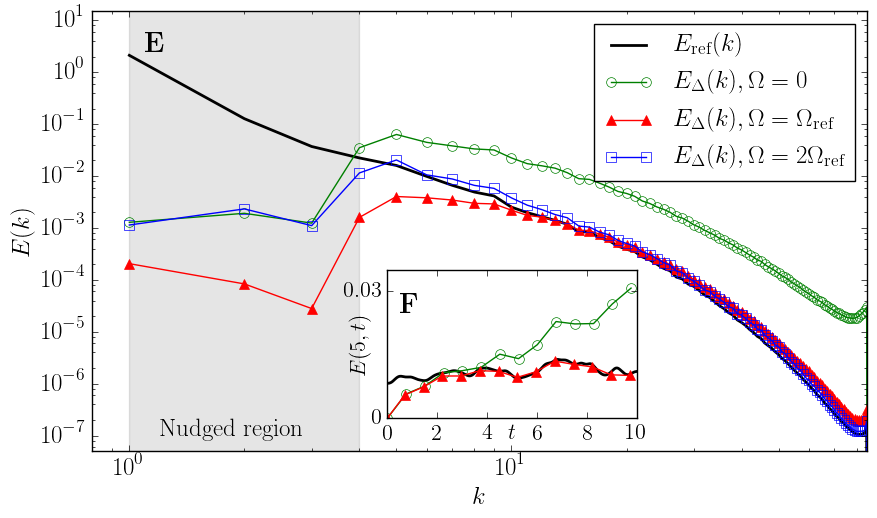}
    \includegraphics[width=0.44\textwidth]{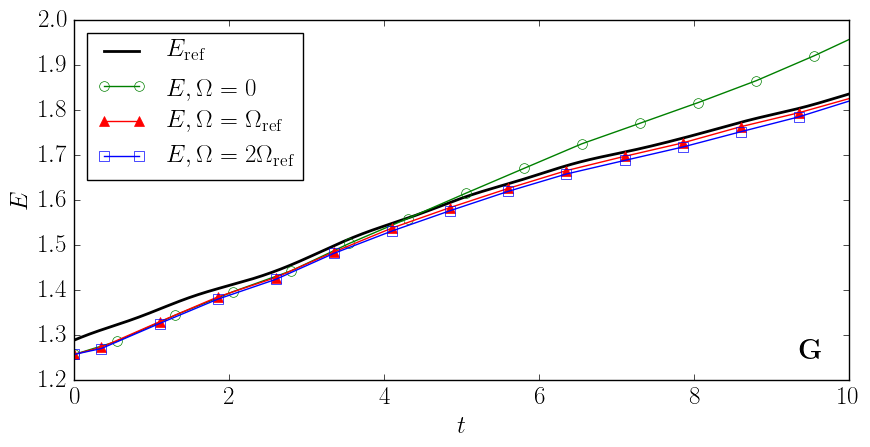}
    \caption{Nudging with different rotation rates. Simulations from
    set-up {\it INFER1} (see Table~\ref{table1}). {\bf A-D}: 2D slices
    of the vorticity field, $\bm{\omega} = \nabla \times {\bf u}$,  in
    the direction parallel to the rotation axis for the reference
    simulation with a rotation frequency $\Oref$, and three nudged
    simulations performed with $\Omega=0, \Oref,$ and $2\Oref$,
    respectively.  {\bf E}: Energy spectra of the reference simulation
    compared with error spectra $E_\Delta(k)$ (see Eq.~\eqref{edelta})
    for different values of the rotation frequency $\Omega$. All spectra
    were computed at the same instants of time. The shaded gray area
    indicate the modes where the nudging is acting.  {\bf F:} Time
    evolution of $E(k,t)$ for $k=5$ for $\Omega=\Oref$ and $\Omega=0$,
    compared to the reference data.  {{\bf G:} Evolution of total energy
    for the reference field and for the three nudged simulations at
    changing $\Omega$.}
    }
    \label{spectra_om}
\end{figure}
%%%%%%%%%%%%%%%%%%%%%%%%%%%%%%%%%%%%5

We start by asking how to guess the exact value of the rotation rate,
$\Oref$, without any {\it a priori} knowledge on its value.  To give a
first idea on the applications of nudging, in panels (A-D) of
Fig.~\ref{spectra_om} we show a series of  2D slices of the vorticity
field in the direction parallel to the rotation axis for the reference
simulation (panel A)  and for three different nudged simulations (panels
B-C-D), two with wrong rotation rates, $\Omega=0$ and  $\Omega=2 \Oref$,
and one with the correct value, $\Omega=\Oref$.  Furthermore, in this
set of simulations we took  ${\cal S}=0$, i.e. we suppose to not know
the forcing mechanism (all simulations are from set-up {\it INFER1}
shown in Table~\ref{table1}).  All  snapshots were taken at the same
instant in time. Comparing the four panels, it is clear that the
simulation nudged with the correct rotation rate (panel C) does
reconstruct the reference flow (panel A) much better then the other two
(panels B and D). \ADDB{It is also worth pointing out that the standard
deviation of the vorticity fields is recovered when rotation is present,
with the values begin around $2.8$ for the reference and the simulations
of both panels C and D, but this is not the case in the absence of
ration (panel C), where the standard deviation takes a value around
$5.4$. All fields have zero mean by construction.} These qualitative
results already provide a first glance  of the two main points we make:
(i) spectral nudging does work well also for fully turbulent 3D flows,
as it does reproduce non-trivial features with high accuracy and (ii) by
optimizing the reconstruction properties, one can infer the unknown
flow-parameters of the nudging flow.  It is worth noticing that the
percentage  of nudged modes is very small, of the order of $\#_{nudged}
\sim 1 \times 10^{-4}$, as we are nudging up to $k=4$ while the maximum
possible wavenumber in this simulation is $k=85$. The nudged modes are
the ones containing the largest amount of energy, but the flow is not
completely determined by their evolution, as many more scales should be
controlled in order to achieve this \cite{Lalescu13}. This fact is clear
when looking at the error spectra in Fig.~\ref{spectra_om}, the error in
the unnudged scales is of the order of the energy at that scales even
though the large scale reconstruction is very good, meaning the unnudged
scales are not slaved to the energy containing modes. Some
synchronization of the small scales is nonetheless present, specially
for the case with $\Omega=\Oref$.  Understanding how much one needs to
nudge in order to fully control a turbulent flow is an open question
that will be addressed in future work.

In order to control the performance of the nudging protocol in
quantitative terms and scale by scale, we introduce a field given by the
difference among the exact input and the one reconstructed  via
\eqref{nsnudged}, $\Delta \bu = \bu -\bu_{\mathrm{ref}}$, and we study
its spectral properties:
\begin{equation}
    E_\Delta(k,t) = \frac12 \sum_{k \le |\bk|<k+1} |\hat{{\bm
    u}}(\bk,t)-\hat{{\bm u}}_\mathrm{ref}(\bk,t)|^2.
    \label{edelta}
\end{equation}
Clearly, the smaller the spectrum $E_\Delta(k)$, the better the
reconstruction. This spectrum will be referred to as the error
spectrum.

In the bottom panel (E) of  Fig.~\ref{spectra_om} we show three
different curves for $E_\Delta(k,t)$ obtained by averaging over all
times when we provide the information, $t_n$, and for the three
different values of the rotation rate,
$\Omega=0,\Omega_{\mathrm{ref}},2\Omega_{\mathrm{ref}}$ already
discussed in panels (A-D), together with the spectrum of the reference
field $E_{\mathrm{ref}}(k) = \sum_{k \le |\bk|<k+1} |\hat{{\bm
u}}_\mathrm{ref}(\bk,t)|^2$, averaged on the same set of times.  In the
figure, the set of nudged wavenumbers is denoted by the grey area.  From
panel (E) it is clear that the optimal nudging is obtained when $\Omega
= \Omega_{\mathrm{ref}}$ is used in \eqref{nsnudged}, as revealed from
the scale-by-scale nudging error, $E_\Delta(k)$, that becomes much
smaller than $E_{\mathrm{ref}}(k)$  for $k \in (k_0,k_1)$. In all
cases, there is a dip in the error spectra at $k=3$, as this is the
first scale at which the forcing is not present in the reference flow,
so the nudging is able to do a better job reconstructing the data. At
$k=4$ the error spectra increases again, mainly because some unnudged
modes are integrated when calculating the spectra at this wavenumber. For
$\Omega=\Omega_{\mathrm{ref}}$, the scale-by-scale error stays smaller
than the reference spectrum up to $k \sim 10$  suggesting a good ability
for data assimilation outside the set of nudged degrees of freedom also.
This latter fact is also confirmed by the inset (panel F)  where we
show the temporal evolution of $E_{\mathrm{ref}}(k,t)$ for an unnudged
wavenumber, $k=5$,  compared with the spectra of the reconstructed field
evolved  with $\Omega=0$ and $\Omega=\Oref$. In this experiment
we started the nudged simulations from zero velocity. As one can see,
after a short transient, only the field evolved with the correct
$\Omega$ rate is indeed able to synchronize with  the time evolution of
the inputting data. Finally, we also show the evolution of the
total energy in panel G for the same simulations. While the case with
$\Omega=0$ is easy to pick apart, the other two are very close to tell
which is one produces a better reconstruction of the flow. This
indicates that comparing averaged quantities (such as the total energy)
may not be the most precise way to determine the value of a parameter.

\begin{figure}
    \centering
    \includegraphics[width=8.5cm]{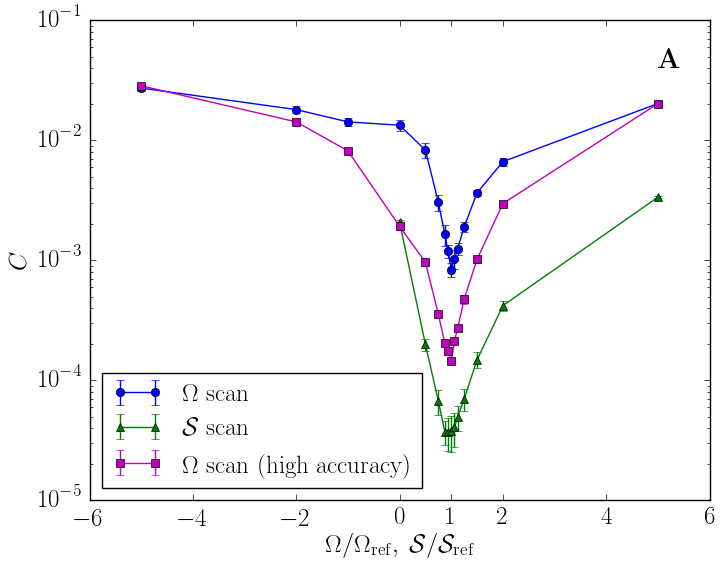}
    \includegraphics[width=8.5cm]{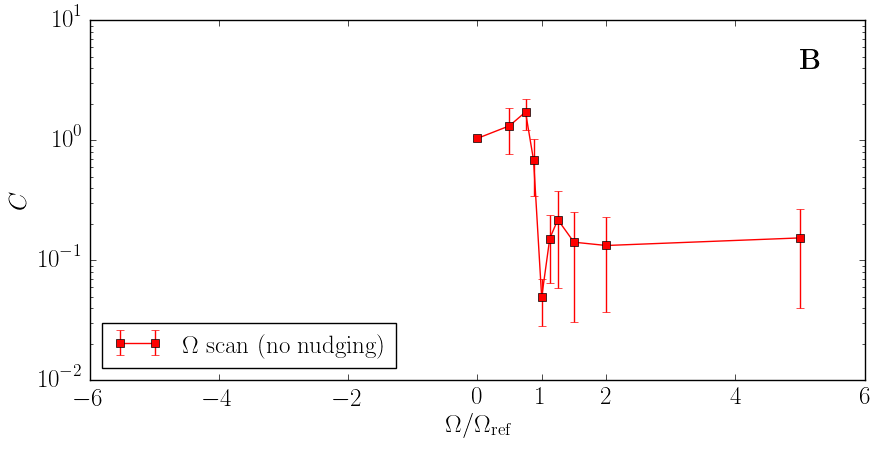}
    \caption{{\bf A}: Value of the mean error committed to reconstruct
    the reference field in the nudged window, ${C}$, for two different
    scanning of the phase space parameters. Blue circle: the case with
    fixed stirring mechanism and at changing the rotation rate $\Omega$
    (set-up {\it INFER1} in Table~\ref{table1}). Green triangles: the
    case with fixed rotation rate and at changing the intensity of the
    stirring parameter, $\mathcal{S}$ (set-up {\it INFER2} in
    Table~\ref{table1}). Magenta squares: a further scan for $\Omega$
    following {\it INFER1} but nudging all wavenumbers up to $k=10$. In
    all cases a clear deep is measured only when the scanning values do
    correspond to the ones used for the reference data, $\Oref$ and
    $\Sref$ respectively. Error bars for each data point were calculated
    by measuring the standard deviation of $C$.  {{\bf B}: Scan of
    $\Omega$ perfomed without nudging but adding the forcing term (i.e.,
    same as {\it INFER1} but with $\alpha=0$ and $\mathcal{S}=\Sref$).}}
    \label{errorscan}
\end{figure}

To be more quantitative about the sensitivity to infer the unknown
rotation rate, we have performed also a detailed scan of $\Omega$ values
around $\Oref$. In Fig.~\ref{errorscan}A we show the
performance of the nudging reconstruction by plotting the value of the
spectrum $E_{\Delta}(k)$, as a function of $\Omega$ and  averaged in
time and in the nudged window:
\begin{equation}
C = \frac{1}{N\,K} \sum_{n=1}^N \int_{k_0}^{k_1} dk E_\Delta(k,t_n),
    \label{error}
\end{equation}
where  $t_n$  are the instant in times where we have measurements and $K
= \int_{k_0}^{k_1} dk E_{\mathrm{ref}}(k)$ is a normalization factor.
Notice that $C$ is defined using  information of the nudging  data only,
i.e. the filtered reference field at the specific times when the
information is provided. In contrast, $E_\Delta(k)$ needs the whole
$\uref$ which in most practical applications would not be available, but
that we can nevertheless access in our numerical experiment.

%%%%%%%%%%%%%%%%%%%%%%%%%%%%%%%%%%%%%%%%%%%%%%%%%%%%%%%%%%%%%%%%%%%%
%%%%%%%%%%%%%%%%%%%%%%%%%%%%%%%%%%%%%%%%%%%%%%%%%%%%%%%

%\begin{equation}
%    \overline{C} = \frac1N \sum_{t_i} C(t_i)
%    \label{avgerror}
%\end{equation}
%with $C$ integrated in the whole nudged window for a set of simulations
%with a refined scanning  of $\Omega$ in an  interval around

From Fig.~\ref{errorscan}A, it is clear the existence of a minimum in
the error when evolving \eqref{nsnudged} with $\Omega \sim
\Omega_{\mathrm{ref}}$. Furthermore, we can determine the correct value
of $\Omega$ with a $6.25\%$ error. The error is calculated by looking at
which values the errorbars for $C$ overlap. We performed another
experiment (set-up {\it INFER2} in Table~\ref{table1}) to test if the
intensity $\mathcal{S}$ of mechanical forcing of the reference
simulation could also be discovered with our nudging protocol. In this
experiment a new reference simulation with $\Omega_{\mathrm{ref}}=0$ and
$\Sref=0.02$ was produced and used to extract the nudging fields (see
Table~\ref{table1} for details). In Fig.~\ref{errorscan}A we show that
the protocol is able to infer the intensity of the stirring mechanism
also, with a  clear minimum of the error \eqref{error} in the proximity
of ${\cal S} \sim {\cal S}_{\mathrm{ref}}$. In this case, the correct
value of ${\cal S}$ can be pinpointed with a $12.5\%$ error.  A third
experiment, following {\it INFER1} but nudging more wavenumbers (so
using more information from the reference as well is shown. Here all
wavenumbers up to $k=10$ where nudged. By doing this we can reduce the
error in the estimation of $\Omega$ to $3.125\%$. All numerical
experiments show that spectral nudging can be used in a physics-informed
way to fit parameters to data and, thus, extract information from it.
\ADDB{Furthermore, in set-up {\it INFER1}, where no information about
the external stirring mechanism is used, performing a one-dimensional
scan (i.e.  varying only the rotation rate) works well. Having said
this, we cannot conclude that this must be the case for generic search
in a multi-dimensional phase-space, where the only systematic way to
proceed would be to adopt a local gradient-descent algorithm.}

A similar scan was performed for the rotation rate but without
using the nudging (i.e., $\alpha=0$). In this case, the forcing term was
also added (with $\mathcal{S}=\mathcal{S}_{\mathrm{ref}}$), otherwise
there would be no energy injection mechanism present. All other
parameters are the same as set-up {\it INFER1}. The results are shown in
Fig.~\ref{errorscan}B. \ADDB{It is clear that obtaining an accurate
value of $\Omega$ out of this scan is very difficult because even though
a minimum is readily seen, the errorbars of several datapoints close to
it overlap. So while running simulations with different parameter values
and performing posterior analysis in order to infer the desired
information is possible, our results suggest the using nudging greatly
improves the sensitivity and accuracy of the search.}

%%%%%%%%%%%%%%%%%%%%%%%%%%%%%%%%%%%%%%%%%%%%%%%%%%%%%
\begin{figure}
    \centering
    \includegraphics[width=8.5cm]{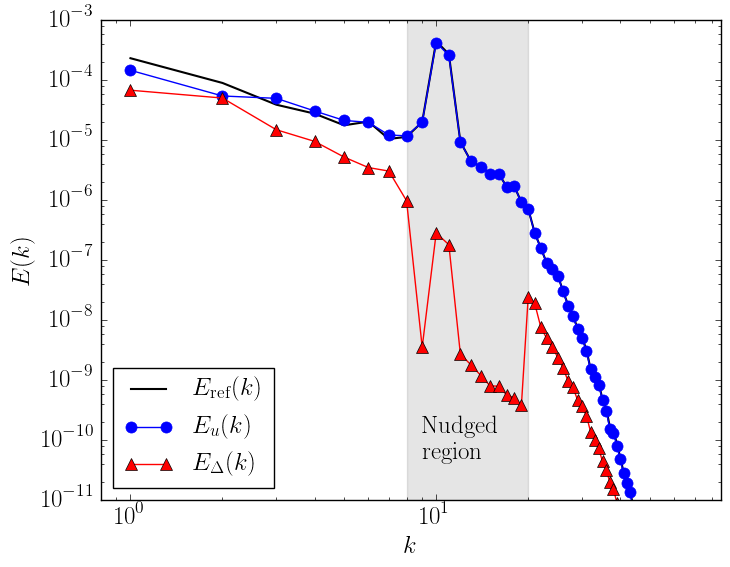}
    \caption{Nudging for the case of rotating turbulence in the inverse
    energy cascade regime. Simulations from set-up {\it PHYS1} (see
    Table~\ref{table1}).
    %an experiment performed using a rotating turbulence simulation
    %forced at $k=10$ with $\Sref=0.004$, and $\Oref=20$ as the
    %reference simulation. This scenario generates an {\it inverse}
    %cascade of energy where energy flows from the smaller to the larger
    %scales, contrary to what usually happens in homogeneous isotropic
    %turbulence. 
    The nudged window is given by the grey area between  $k_0=8$ and
    $k_1=20$.
    %with $\alpha=10$ and $\tau=0.01$. 
    The reference spectrum $E_{\mathrm{ref}}$ and the nudged spectrum
    $E_u$ almost coincide for $k>8$, making it hard to discern between
    the two.
    Both the intensity of the forcing $\mathcal{S}$ and the rotation
    rate $\Omega$ are zero in the nudged simulation, so all energy
    injection and anisotropic effects are coming from the nudging term.
    Notice the strongly reduced error spectrum $E_\Delta(k)$ for a large
    set of wavenumbers, indicating an optimal reconstruction quality.}
    \label{invcasc}
\end{figure}
%%%%%%%%%%%%%%%%%%%%%%%%%%%%%%%%%%%%%%%

\begin{figure}
    \centering
    \includegraphics[width=8.5cm]{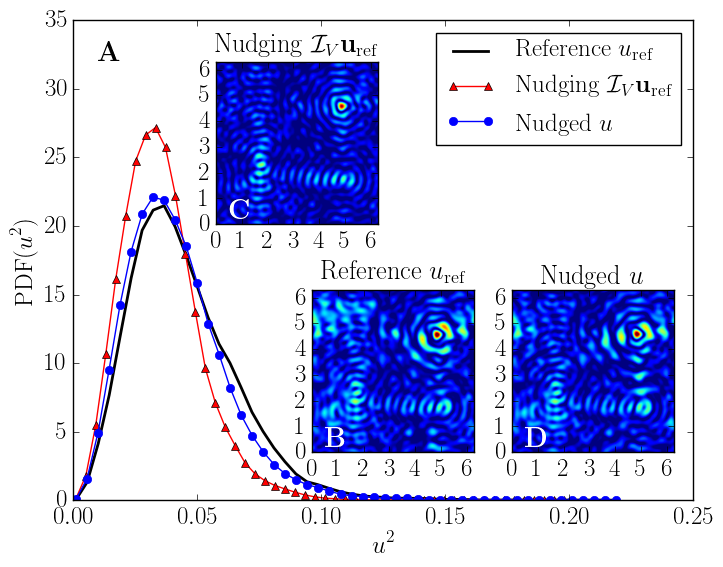}
    \caption{{\bf A}: Probability density functions (PDF) of the point-wise kinetic energy for the reference simulation $|\uref({\bf x})|^2$ (continuous black
    line), the nudged simulation $|\bu ({\bf x})|^2$ (circles) and the nudging
    input field $|\mathcal{I}_V \uref({\bf x})|^2$ (triangles) for the inverse
    cascade experiment (set-up {\it PHYS1} in Table~\ref{table1}). {\bf
    B-D}: 2D slices of planes perpendicular to the rotation axis of the
    absolute velocity fields.}
    \label{invcasc_pdf}
\end{figure}
%%%%%%%%%%%%%%%%%%%%%%%%%%%%%%%%%%%%%%%%%%%%%%%%%%%%%%%%%%%%%%%%%%%%%%%

\section*{Inferring the large-scale velocity distribution without input rotation}

In this section we describe how to use nudging to infer,  under some
circumstances, the entire set of  large-scale physical flow structures
of the reference data without a detailed knowledge of the forces acting
on flow.  To test this idea we performed a new experiment by using a
turbulent flow under rotation and in the presence of an inverse energy
cascade. It is well known that if rotation is strong enough and energy
is injected at large wavenumbers the flow undergoes a transition from a
direct to a split turbulent energy cascade, accumulating kinetic energy
and producing a non-trivial cyclonic distribution of vortices at larger
and larger scales \cite{Smith96,Mininni09,Mininni09b}. This regime does
not occur naturally in homogeneous isotropic three-dimensional
turbulence \cite{Biferale13}, but it is argued to be important in
many geophysical set-ups  in the oceans \cite{Scott05,Corrado17} and in
the atmosphere \cite{Lacorata04}. Here, we  show how a suitable nudging
strategy is indeed able to reconstruct the inverse energy cascade even
in the absence of any explicit rotation term in the nudged equations
\eqref{nsnudged}, provided that the $\uref$ is inputting information
around the injection scale. To do this we use a rotating turbulent flow
forced at $k_f=10$ and with $\Omega_{\mathrm{ref}} =20$ and
$\Sref=0.004$ as a reference (set-up {\it PHYS1} in Table~\ref{table1})
where an inverse energy cascade develops. We then evolve
\eqref{nsnudged} without any rotation and any external forcing: $${\bf
{\cal F}}[{\bf u}_{\mathrm{ref}}, {\cal V}_{\mathrm{ref}}]=0$$  in this way we
are completely ignorant about the physics we want to reproduce.   In
Fig.~\ref{invcasc} we show  that by  nudging in the region around the
injection mechanism,  the energy spectra of the reference simulation is
well reproduced by  the nudged simulation even, and in particular, in
the inverse energy cascade range. The presence of a strong peak
around the forced wavenumber is typical of systems where an inverse
cascade is present, as this is a slow and inefficient transfer mechanism
\cite{Smith96,Mininni09,Mininni09b,Biferale16}. Even though the only information we
input is the nudging filtered field, the nudged evolution is able to
reconstruct the inverse cascade and the correct spectrum slope even for
scales much smaller then the ones where we nudge. It is remarkable how
the spectrum error, $E_\Delta(k)$ is small also for modes outside the
nudging window $k<k_0$ and $k>k_1$, indicating the presence of strong
non-local spectral correlation in the split-energy cascade mechanism
which are fully reconstructed by our protocol.

To go beyond spectral properties and to check the ability to reconstruct
the large-scale coherent structures in the rotating flow,  we plot in
Fig.~\ref{invcasc_pdf} the probability density functions (PDFs) of the
space-dependent kinetic energy  for the  reference simulation, $
|\uref|^2/2$,  the nudged simulation, $ |\bu|^2/2$, and the nudging
field, $|\mathcal{I}_V \uref|^2/2$. As one can see, the reconstructed
field has a PDF very close to the reference case, even if the nudging
input field does not. In the same figure, we also show  2D slices of the
absolute velocity fields in planes perpendicular to the rotation axis
for the three fields as before. As one can see, the nudged simulation
(panel D) is able to extrapolate the unknown large-scale reference flow
structures  extremely well (panel B), for a case where the nudged
inputting data do not contain any information about those scale (C).
The apparent patterns seen in these visualizations are a product
of the strong forcing present in the system acting around $k=10$.

%This experiment then shows how the nudging protocol
%%can not only be used to infer parameters but also to infer physics from
%data and how it can be used to answer concrete and current questions
%present in the earth sciences, such as if the oceans or the atmosphere
%present inverse cascades.

\section*{Conclusions}

Spectral nudging is a physics-informed  technique commonly used to guide
the evolution of chaotic dynamical systems inputting measured data.
Giving examples for both isotropic and rotating 3D turbulence, we have
shown how this technique can  be efficiently used to infer both the
physical parameters entering in the external stirring forces and the
large-scale velocity distribution for the inverse energy cascade regime,
typical of strongly rotating turbulent flows. The method can be further
improved and optimised by using different nudging parameters for
different degrees of freedoms, e.g.  by changing $\alpha$ and $\tau$
with ${\bf k}$. A detailed study of nudging performances for homogeneous
and isotropic turbulence at different Reynolds numbers, different
nudging windows and at changing the spatial locations of the
measurements stations will be reported elsewhere.

% The method  is also efficient and economic to operate compared to other
% strategies used for parameter estimations which make use of
% supplementary differential equations (e.g. the adjoint evolution
% equation) to construct the gradient of a suitable cost function
% appearing in the variational problem \cite{Navon98}. 

Other strategies used to estimate parameters, such as variational
methods \cite{Navon98} or ensemble based methods \cite{Anderson99,
Anderson01, Ruiz13}, require the need to postulate and error correlation
matrix and make assumptions about the behavior of the errors and
deviations, need to use linearized models (for variational methods), or
are based on minimizing complicated functions (again for variational
methods). Nudging based strategies do require to perform several forwards
simulations, similar to ensemble based method. One advantage other
methods have compared to nudging, is the ease to incorporate information
on observables (such as precipitation, for example) and not just state
variables (such as the velocity field, as was used here. Interestingly,
variational data assimilation schemes have been exploited to determine
vectors of optimal nudging coefficients \cite{Zou92}.  Here, we reversed
the point of view: given the coefficients $\alpha, \tau$, we employed
nudging to estimate the physical flow parameters. Finally, the method is
also general and extendable to other problems, opening the route to
applications for parameter inferring to a vast set of hydrodynamical
situations including, to cite just the most promising cases, i)
optimising sub-grid-scale models in Large Eddy Simulations, by
inferring parameters against data extracted from either observation or
benchmark direct numerical simulations; ii)
large-scale turbulent transport to determine eddy-viscosity  and
eddy-diffusivity \cite{Yu91,Lin01}; iii) the identification of ambient
air  sources and the quantification of their contribution to pollution
levels (the so-called source apportionment problem) \cite{Bove14}; iv)
partial field reconstruction using advanced Lidar systems
\cite{Cooper97} to reveal the free parameters characterizing the
atmospheric boundary layer; v) correction of velocity fields in ocean
circulation models with Lagrangian data (e.g. from drifting buoys)
\cite{Taillandier06a,Taillandier06b} and/or other sources including HF
radar data \cite{Berta14}.

\begin{acknowledgments}
The authors acknowledge funding from the European Research Council under
the European Community's Seventh Framework Program, ERC Grant Agreement
No.~339032.
\end{acknowledgments} 

\bibliography{ms}

\end{document}